\documentclass[aps,twocolumn,floats,nofootinbib]{revtex4} 
\usepackage{graphics,graphicx,epsfig} 
\usepackage{amssymb,color} 
\usepackage{epsf,epstopdf,wrapfig} 
\usepackage {amsmath} 
 \usepackage{textgreek} 
\usepackage{sidecap}

\newcommand{\beq}{\begin{equation}} 
\newcommand{\eeq}{\end{equation}} 
\newcommand{\beqn}{\begin{eqnarray}} 
\newcommand{\eeqn}{\end{eqnarray}}

\newcommand {\pa}{\partial}

\newcommand{\lan}{\langle}
\newcommand{\ran}{\rangle}
\newcommand {\pnl}{\left}
\newcommand {\pnr}{\right}

\newcommand {\te}{\text}

\newcommand{\imp}{\implies}
\newcommand{\ift}{\infty}

\newcommand{\fr}{\frac}

\newcommand{\dd}[2]{\fr{\te{d}#1}{\te{d}#2}}
\newcommand{\ddd}[2]{\fr{\te{d}^2#1}{\te{d}#2^2}}
\newcommand{\pp}[2]{\fr{\pa #1}{\pa #2}}

  \newcommand{\pn}[1]{\pnl ( #1 \pnr )}

\newcommand{\expec}[1]{\pnl \lan #1 \pnr \ran}

\begin{document}

\title{The statistical mechanics of Twitter} 


\author{Gavin Hall$^{a,b}$ and William Bialek$^{b,c}$}
\affiliation{ $^a$Feinberg School of Medicine,  Northwestern University, Chicago, IL 60611\\
 $^b$Joseph Henry Laboratories of Physics, Princeton University, Princeton, NJ 08544\\
 $^c$Initiative for the Theoretical Sciences, The Graduate Center, City University of New York, 365 Fifth Ave., New York, NY 10016}

\date{\today} 

\begin{abstract}  
We build models for the distribution of  social states in Twitter communities.  States can be defined by the participation vs silence of individuals in conversations that surround key words, and we approximate the joint distribution of these binary variables using the maximum entropy principle, finding the least structured models that match the mean probability of individuals tweeting and their pairwise correlations.  These models provide very accurate, quantitative descriptions of higher order structure in these social networks.  The parameters of these models seem poised close to critical surfaces in the space of possible models, and we observe scaling behavior of the data under coarse--graining.  These results suggest that simple models, grounded in statistical physics, may provide a useful point of view on the  larger data sets now emerging from complex social systems.
\end{abstract}

\pacs{Valid PACS appear here} 

\maketitle

\section{\label{sec:level1}Introduction}

Social systems exhibit rich collective behaviors. Many large-scale social processes, from cultural fads \cite{Fad} to residential segregation \cite{Schelling} to the polarization of political opinions \cite{Polarization}, depend on the interactions of many individuals. Indeed, many social phenomena are emergent almost by definition. Sociologists have long explored the relationship between individual actions, interactions among individuals, and macroscopic social outcomes \cite{Brodbeck,Sawyer}. 

While there is general agreement on the qualitative idea that social phenomena are emergent, there has been much less progress toward quantitative theories.  For inanimate systems, especially near thermal equilibrium, statistical mechanics  provides a framework for building quantitative theories of how  macroscopic behaviors emerge from microscopic interactions.  Importantly, successful theories in statistical mechanics often are simpler than the underlying microscopic reality, and the renormalization group allows us to understand how this simplification is possible \cite{Sethna, ManyLength}.  Inspired by these examples, there have been efforts to  examine social phenomena using ideas and methods borrowed from statistical physics \cite{SocRev}. Examples include the dynamics of a strike \cite{IsingOpinion}, the emergence of group consensus \cite{SznajdRev,voter}, the behavior of dancers at heavy metal concerts \cite{Mosh}, and temporal patterns of activity and inactivity on Twitter \cite{TwitterBot}.
 
In much previous work, methods from statistical physics were used to construct mathematically precise versions of existing sociological theories   \cite{Bruch}, but the resulting models might or might not engage with quantitative data on real social systems.  Here, inspired by a stream of work on biological systems ranging from families of proteins \cite{bialek+ranganathan_07,weigt_al_09,antibody,marks+al_11} to networks of neurons  \cite{WeakPairwise, IsingRealNeurons, NeuronCriticality, CollectiveNeurons} to flocks of birds  \cite{Birds, Birds2}, we take a different approach, using the maximum entropy method \cite{Jaynes,Jaynes2} to build a statistical mechanics description of a social system  directly from real data, independent of traditional sociological hypotheses.  Previous efforts in this direction include analyses of voting patterns on the US Supreme Court \cite{supreme, Lee2} and patterns of conflict in troops of macaques \cite{macaque}.

Here we adopt the strategy of building models directly from data, and explore the emergence of collective behaviors in Twitter communities.  In order to carry out this program we need to identify communities and to define behavioral states for all the individuals in those communities.  As a  first step, we take these states to be participating or not participating in conversations that involve particular keywords. Then, states are binary and the maximum entropy models consistent with the pairwise correlations among these variables are equivalent to Ising spin glasses \cite{MezardSpinGlass}.   These relatively simple models successfully predict higher order structure in the data.  Analysis of these models, as well as a direct coarse--graining of the data, suggests that these systems are close to a critical point or critical surface in their parameter space. We explore what this might mean for social functioning. 

\section{Networks and states}

The full set of Twitter users is vast, beyond our ability to build explicit models.  As a start, we want to focus on smaller networks of users who are well connected with one another.  We start by choosing a single Twitter user and then find the people whom this user follows, and the people whom those neighbors follow.  The result is a social network with known connectivity and relatively short path lengths.  Twitter provides public access to the last 3200 tweets for each user, so each node in our network is associated with a stream of timestamped text.  

We note that the initial choice of root users is arbitrary, and it is difficult to ask in what sense our results are representative (except by trying many examples).  Happily, none of the individuals identified in this way were public figures, or otherwise strong outliers in terms of their social media presence.  

\begin{figure}[bt]
  \includegraphics[width=\linewidth]{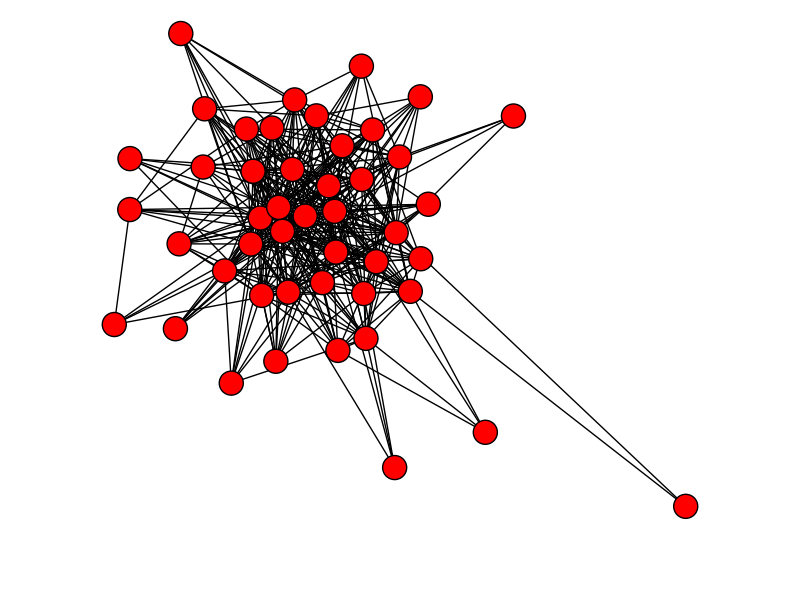} 
\caption{ \textbf{Example Social Network.} An example social network used for building a pairwise max-ent model. Same sub--community as used in Fig 4. We include more information on the topology of the social networks examined in Appendix A.
\label{network}}
\end{figure}

Even the networks at depth two from a random user are quite large, so we focus further,  breaking these networks into sub--communities using the Clauset--Newman--Moore algorithm  \cite{community}.  This algorithm builds sub--communities such that the proportion of edges within sub--communities is maximized while minimizing the number of edges between sub--communities. The resulting sub--communities are the networks that we use for further analysis.  Among 106 examples we analyze networks that contain between 8 and 80 people, and there don't appear to be any simple trends of topology vs size (see Appendix A for these and other details).  An example of the networks that we identify is shown in Fig \ref{network}.

Having defined a network of individuals ${i} = 1, 2,\cdots , N$, what are the states $\sigma_i$ taken on by these individuals?  In examining the raw data, we find prominent words that are used many times within a short period of time (Appendix B).  For the remainder of this paper, we will call these prominent words `keywords,' and they can intuitively be thought of as something akin to a topic of conversation in the community being studied. 
For example, in a community with many physicists, one prominent keyword identified was `Kosterlitz', as many people talked about J.M. Kosterlitz in a very short period of time after he shared the 2016 Nobel Prize. 

These keywords suggest a simple and intuitive way to binarize data from Twitter:   either a given individual has used a given keyword, or they have not. So, in the physicists' example, we can find who talked about Kosterlitz and who did not, and assign everyone who did talk about Kosterlitz a state $\sigma _i = 1$ and everyone who did not talk about Kosterlitz a variable $\sigma _i = -1$. We can then conglomerate these variables into a vector $\boldsymbol \sigma$ representing whether or not everyone in the dataset used the given keyword.   In this sense, the variable $\boldsymbol \sigma$ represents a social state of the community---there is an event happening in the world (represented by the keyword), and   members of the community can either participate in this event or remain silent.   

The succession of keywords, which by definition are each well localized in time,  provide a series of snapshots of the social state $\boldsymbol \sigma$.  These snapshots are drawn out of some distribution $P( \boldsymbol \sigma )$ which characterizes the collective states taken on by the network as a whole.  Our goal is to characterize this distribution.

There obviously are many ways  to simplify or binarize data from Twitter. Even with the use of keywords as a tool for simplification, these keywords themselves could be chosen in different ways.   While we cannot claim uniqueness, we do feel that our choice of simplification is intuitive and easy to implement (Appendix B).  Importantly, we will see that the states defined in this way have orderly behavior.

\section{Maximum entropy models}

The social states $\boldsymbol \sigma$ are the ``microscopic'' states of our system, describing what each individual user is doing during a single conversation.  In the spirit of statistical mechanics, we would like to write down the analog of the Boltzmann distribution, $P(\boldsymbol \sigma )$, which tells us   which social states are favored in a community and which states are disfavored.  As usual, the number of possible states $\boldsymbol \sigma$ is so large ($2^N$) that we cannot directly ``measure'' $P(\boldsymbol \sigma )$ from any reasonable amount of data once we are looking at networks of reasonable size ($N\gg 10$).  More precisely, the distribution $P(\boldsymbol \sigma )$ is a list of length $2^N$, constrained only by normalization, and so if there is no simpler underlying structure then we can't make any progress without making more than $2^N$ measurements.

The maximum entropy method, as first emphasized by Jaynes  \cite{Jaynes, Jaynes2}, gives us a way of searching systematically for simplified descriptions.   Without constraints, the distribution that maximizes the entropy, and hence has the minimal structure, is the uniform distribution, which is a clearly unrealistic model. We therefore introduce constraints to ensure that the model is capable of reproducing features of the empirical data, and then maximize the entropy to remove any structures which are not absolutely necessary to meet these constraints.

\begin{figure}
\includegraphics[width=\linewidth]{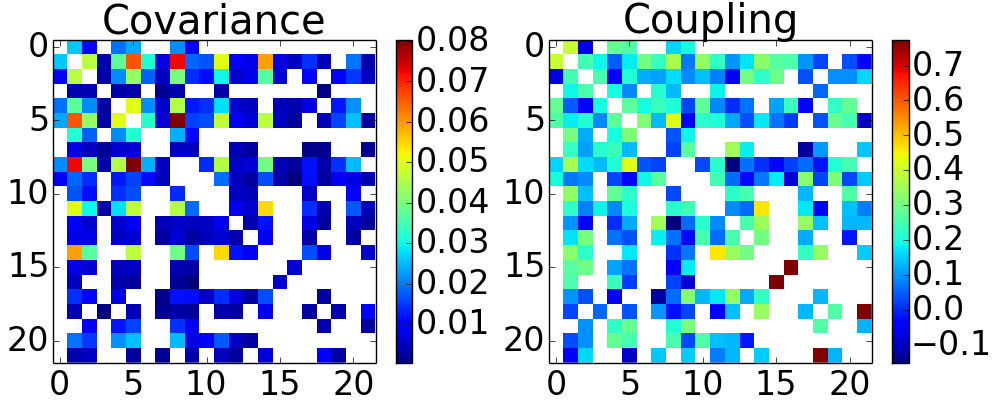} 
\caption{ \textbf{Covariance and coupling matrices.} An example of the covariance matrix [Eq (\ref{Cij_def})] and the corresponding coupling matrix $J_{ij}$  [Eq (\ref{E})] for a Twitter community. Same community as used in Figures \ref{triplets} and \ref{Qfig}.  Blank elements correspond to pairs of users without a direct social connection.\label{matrix}}
\end{figure}

In the present context, it  is natural to constrain one--body marginals, which means that our model will reproduce the observed average activity of each user in the community,
\begin{equation}
\langle \sigma_i \rangle_P \equiv \sum_{\boldsymbol \sigma}P(\boldsymbol \sigma )\sigma_i = \langle \sigma_i \rangle_{\rm obs} .
\label{fixh}
\end{equation}
To capture the interactions among individuals, we insist that our model also match the correlations between  pairs of users, so that
\begin{equation}
\langle \sigma_i \sigma_j \rangle_P \equiv \sum_{\boldsymbol \sigma}P(\boldsymbol \sigma )\sigma_i \sigma_j= \langle \sigma_i \sigma_j\rangle_{\rm obs} ;
\label{fixJ}
\end{equation}
to respect the known connectivity of the network, and to limit the complexity of our models, we enforce this constraint only among pairs that have a direct social connection.  Given these constraints, we look for the distribution $P(\boldsymbol \sigma )$ with the largest possible entropy, and the form of this distribution (Appendix C) is
\begin{align}
P(\boldsymbol \sigma)&= \fr{1}{Z} e^{-E_{\boldsymbol\sigma}} \label{boltz}\\
E_{\boldsymbol \sigma} &= -\sum_i h_i\sigma_i  - \fr{1}{2} \sum_{i\neq j} A_{ij}J_{ij} \sigma _i \sigma_j ,\label{E}
\end{align}
where $h_i$ are parameters corresponding to constraints on one body marginals [Eq (\ref{fixh})] and $J_{ij}$ are parameters corresponding to constraints on two body marginals [Eq (\ref{fixJ})].  We introduce the  adjacency matrix $A$ for the community to remind us that we have constraints only among connected pairs; $A_{ij} = 1$ if there is a social tie between individuals $i$ and $j$ and 0 otherwise.  We find the values of $\{h_i, J_{ij}\}$ by solving  Eqs (\ref{fixh}, \ref{fixJ}).  This is in general  a difficult computational problem \cite{InverseIsing}, but we have solved it numerically for communities of up to 80 people using Monte Carlo methods \cite{MCFast} (see Appendix C for details).  
An example of a covariance matrix,
\begin{equation}
C_{ij} =  \langle \sigma_i \sigma_j\rangle_{\rm obs}  -  \langle \sigma_i \rangle_{\rm obs} \langle \sigma_j\rangle_{\rm obs} 
\label{Cij_def}
\end{equation}
and the corresponding coupling matrix $J_{ij}$ for a community is shown in Fig \ref{matrix}. 

\begin{figure}[b]
  \includegraphics[width=\linewidth]{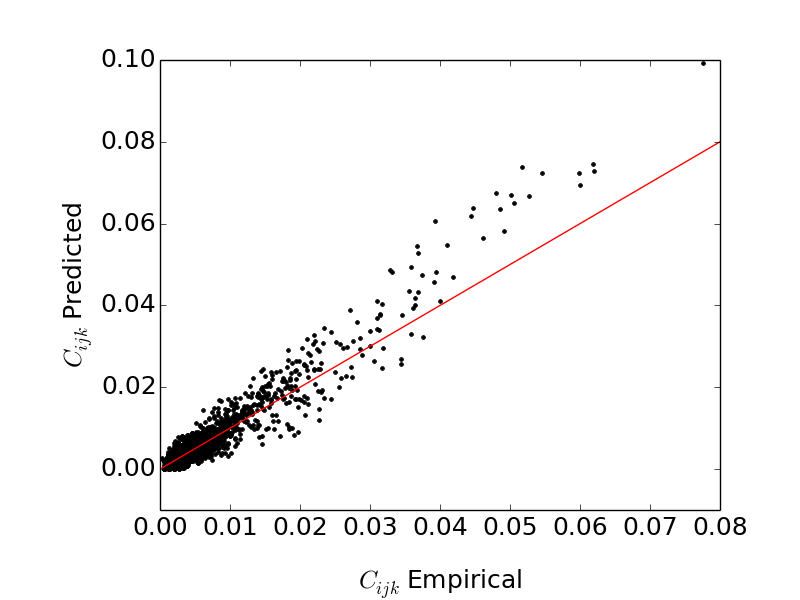} 
\caption{ \textbf{Triplet correlations.} Predicted three point correlations from our model vs the empirical value of the three point correlations estimated from data, for one sub--community.  Error bars typically are $\sim 10\%$ of the measured values; see Appendix E.  \label{triplets}}
\end{figure}

Since real communities exhibit correlations with both signs, the model in Eqs (\ref{boltz}, \ref{E}) corresponds to an Ising spin glass \cite{MezardSpinGlass, EdwardsAnderson}.  Importantly, the couplings $J_{ij}$ are not completely random, but are determined by the observed correlations. Generally, the couplings $J_{ij}$ can take on both positive and negative values, and frustration is common  (Appendix D). 

Maximum entropy models are appealing both because of their simplicity and because of their connections to statistical physics.  But these are not arguments for their correctness.  We could easily imagine, for example, that there are important multibody interactions among individuals, and in this case we could not give an accurate description of the joint distribution by matching pairwise correlations alone.  To test these models we can compute higher order statistical quantities, and ask if these agree with the data.  Importantly, once we have matched the one--body and two--body marginals, there are no free parameters left to adjust, so we are not ``fitting'' these higher order structures---either we get them right or we get them wrong.  We focus here on two such structures, the triplet correlations and the distribution of how many people tweet about each keyword.

For every distinct group of three users in a network, we can define the triplet correlation
\begin{equation}
C_{ijk} = \langle (\sigma_i  - \expec{\sigma_i})(\sigma_j  - \expec{\sigma_j})(\sigma_k  - \expec{\sigma_k}) \rangle .
\end{equation}
These correlations typically are quite small ($C\sim 0.01$) but can be estimated with fractional errors $\sim 10\%$ given the sizes of our data sets (Appendix E).  In Fig \ref{triplets} we show the comparison of predicted vs observed triplet correlations, in one sub--community of 22 users.  Results for many other sub--communities are similar, with prediction errors on the same scale as our measurement errors,  although in certain communities the pairwise model fails to capture some aspects of three point correlation structure (see Fig \ref {tripletError} in Appendix E for examples). 

  \begin{figure}[t]
   \includegraphics[width=80mm]{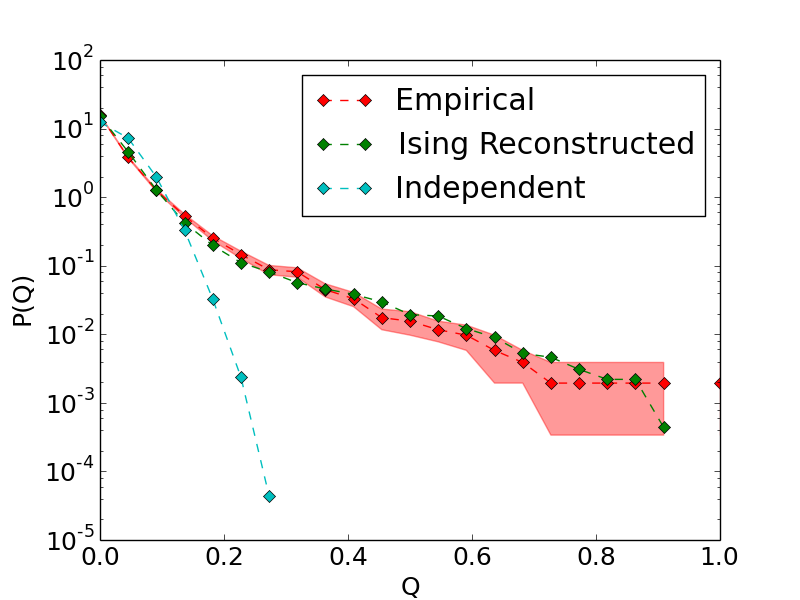} 
   \caption{\textbf{Distribution of simultaneous activity.} Probability that a fraction $Q$ of users tweet about a particular keyword. Empirical data (red), maximum entropy model (green), and a maximum entropy model including only one--body constraints (cyan). \label{Qfig}}
\end{figure}

A different way of assessing higher order structure in the network is to ask about the fraction of users that participate in a conversation,
\begin{equation}
Q = \fr{1}{2}+\fr{1}{2N}\sum _{i=1}^N \sigma _i .
\label{Qdef}
\end{equation}
If the users tweet independently, then for large $N$ we would see a Gaussian distribution of $Q$, and even for smaller communities the tail at large $Q$ would be very restricted.   We see in Fig \ref{Qfig} that the observed distribution of $Q$ is quite broad, with an extended tail, and that this is captured within error bars by our model.  Thus, although we match only correlations among pairs of users, we can predict, quantitatively, the probability that many users will be active in the same conversation. 

There are many reasons why a pairwise maximum entropy model should not work.  In particular, even in the absence of explicit combinatorial effects, averaging over many unseen factors that affect all the users, or different subsets of users, will generate effective multibody interactions in the joint distribution.  These effects may be present, but what we see from Figs \ref{triplets} and \ref{Qfig} is that we don't need to make explicit models of these effects in order to generate quantitative predictions for the joint distribution of social behaviors.   These models, while simple, are sufficiently precise that it makes sense to take them seriously as statistical physics problems and ask what we can learn about the collective behavior of the network.

\section{Toward a phase diagram}

A crucial lesson of statistical physics is that the parameter space of models for systems with many interacting degrees of freedom breaks up, at large $N$, into distinct phases with qualitatively different behaviors.  The boundaries between these phases become sharp as $N\rightarrow\infty$, and on the boundaries the behavior of the system is a singular function of its parameters.  Here we try to locate real networks of Twitter users in relation to these critical surfaces in parameter space.

Maximum entropy models have  the form of a Boltzmann distribution, and so we can think about an `energy landscape' as a function of the social state $\boldsymbol \sigma$; energy minima correspond to probability maxima, identifying states that are favored by the network.  Remarkably, every sub--community we have examined has the same dominant energy basin that contains the vast majority of the data. This basin is defined by silence in the sub--community ($\sigma _i = -1$ for all $i$). This in turn allows for an enormous simplification of our data, as we can define an order parameter by the overlap with the silent state \cite{MezardSpinGlass}.  The overlap with silence is just the negative of the usual magnetization, but it's important that we don't choose the magnetization arbitrarily.   

  \begin{figure}[b]
   \includegraphics[width=\linewidth]{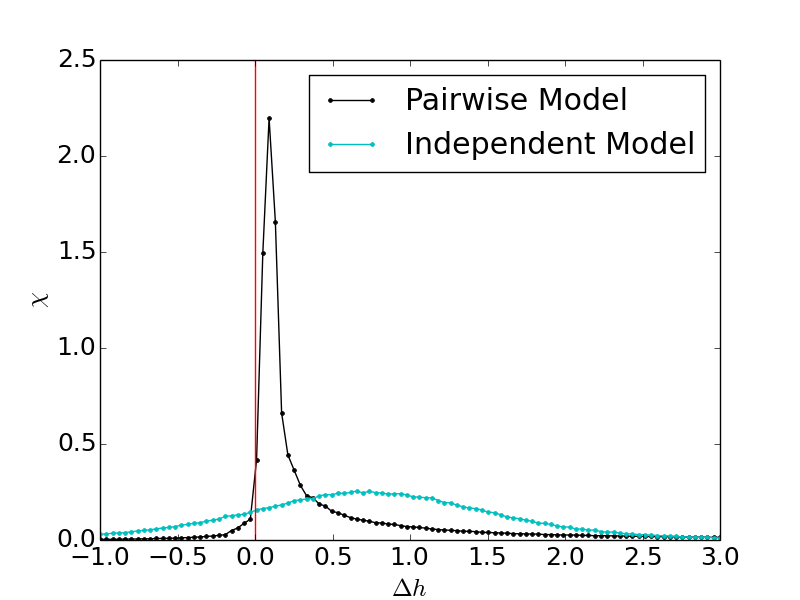} 
   \caption{\textbf{Susceptibility.} Predicted susceptibility against a forcing field for the community of 46 people shown in Fig \ref{network}. Pairwise maximum entropy model (black) and independent model (cyan); red line indicates $\Delta h=0$, corresponding to the parameters inferred for the real network. \label{chi}}
\end{figure}

Once we have an order parameter, we can define, in the usual way, a conjugate field and a susceptibility of the order parameter to this field; the susceptibility will be equal to the variance of the order parameter.  Again this example is relatively simple---the conjugate field is a uniform ``magnetic field'' $\Delta h$ that biases each user to tweet or remain silent, and the susceptibility $\chi$ is proportional to the variance of the fractional activity $Q$ defined above [Eq (\ref{Qdef})].  In Fig \ref{chi}, we track $\chi$ vs $\Delta h$ in a sub--community of 46 people (Fig \ref{network}).

As we can see in Figure \ref{chi}, the system exhibits a large peak in susceptibility at a small forcing field. This is a collective effect, and would not present if the users all tweeted independently (as shown in cyan).    This peak in susceptibility is reminiscent of what we see at a critical point, where incremental changes in the control parameter lead to disproportionately large changes in observable behavior.  Relative to the width of the peak, the system seems to be poised quite close to this  near--critical point. 

One may object that the language of ``applied fields'' involves taking the mapping between maximum entropy models and their physical counterparts a bit too seriously.  As an alternative, we can bias the system by choosing individuals out of the community and conditioning the distribution of all other users on these individuals being in the state $\sigma_\mu = 1$ (tweeting).  Mathematically, if we hold $\sigma_\mu = 1$ for $\mu = \mu_1, \mu_2, \cdots$, then for all the remaining users the joint distribution still is given by Eqs (\ref{boltz}, \ref{E}), but with  $h_i \rightarrow h_i + \sum_\mu J_{i\mu}$ \cite{macaque}. 

The fact that forcing some users to be active will bias the mean activity of other users is not surprising.  More interesting is that the variance of total activity in the other users changes nonmonotonically as a function of the number of users that we force, echoing the dependence of susceptibility on applied field.     In all the sub--communities that we have examined, the peak in variance occurs upon forcing just a handful of users, often just one; there is no indication that this depends systematically on  $N$.    We conclude that many Twitter communities are within one user of being near maximal variance in activity \cite{gavThesis}.  This is a direct but perhaps more intuitive analog of the peak in susceptibility for very small applied fields (Fig \ref{chi}).

If we have a statistical mechanics, then it should be possible to construct a thermodynamics.  Much of thermodynamics is about the tradeoff between energy and entropy, and it might be unclear what this has to do with tweeting.  But in the Boltzmann distribution, energy is just the (negative) log probability, and (microcanonical) entropy counts the number of states that have this probability.  Intuitively, social states in which more users are active have lower probability (Fig \ref{Qfig}), but until fully half the users are active there are more distinct states available at larger $Q$ [Eq (\ref{Qdef})].  Thus,  less probable (higher energy) states are more numerous (higher entropy).   We explore this tradeoff between probability and numerosity following Refs \cite{NeuronCriticality,BialekCriticality,NaturalImages}; for details see Appendix F.

The maximum entropy model assigns to each state $\boldsymbol \sigma$ an energy $E_{\boldsymbol\sigma}$, through Eq (\ref{E}).   We would like to count the number of states that have a particular energy, or range of energies, but this involves making bins along the energy axis.  A simple alternative is to count the number of states with energy less than $E$, so we define the entropy
\begin{equation}
S(E)= \ln \pn {\sum _{\boldsymbol \sigma} \Theta (E - E_{\boldsymbol \sigma})}
\label{mc_ent}
\end{equation}
where $\Theta (x)$ is the step function: $\Theta (x>0) =1$, $\Theta (x<0) =0$.  We recall that the temperature is the derivative of the entropy with respect to energy. In our case we have $T =1$, from Eq (\ref{boltz}), and so the condition
\begin{equation}
{{dS(E)}\over{dE}} = 1
\end{equation}
picks out the typical energy of the system.  The fluctuations around the typical energy are related to the heat capacity $C$,
\begin{equation}
\langle (\delta E)^2\rangle   =  \pn{ -\ddd{S}{E}}^{-1} = C
\end{equation}
again with $T=1$.  We expect that energies and entropies both are extensive, that is proportional to system size $N$ for large $N$, so that $C$ itself also is of order $N$.  Then the fractional fluctuations  $\langle (\delta E)^2\rangle/E^2 \sim 1/N$ vanish rapidly for larger systems.  At many critical points, $d^2S/dE^2$ vanishes, the specific heat $C/N$ diverges with $N$, and the variance of energy fluctuations are similarly large.

  \begin{figure}[t]
   \includegraphics[width=\linewidth]{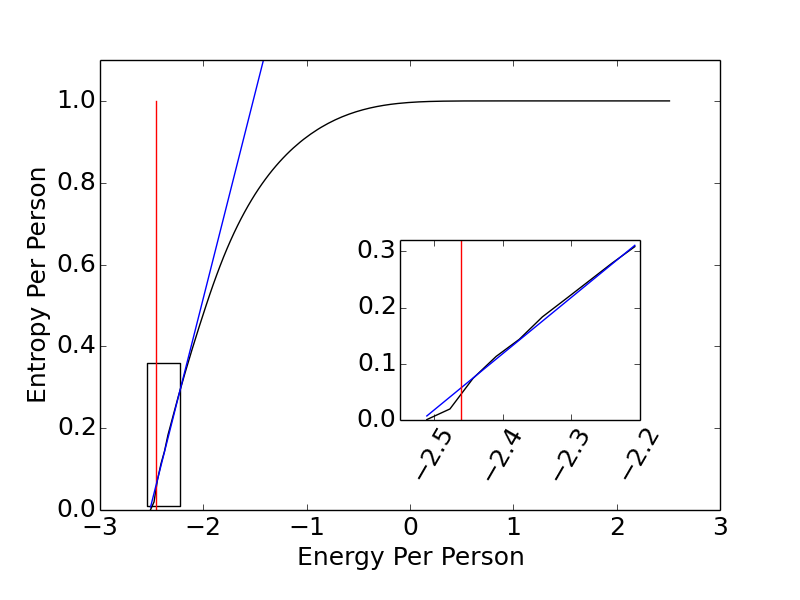} 
   \caption{\textbf{Entropy vs energy.} $S(E)/N$ from Eq (\ref{mc_ent}), plotted vs $E/N$, for a sub--community of $N = 52$ users.   Red line indicates average energy of system, blue line indicates line of slope 1 fit around the actual energy. \label{SvsE}}
\end{figure}

Starting with our maximum entropy model, we can  find entropy as a function of energy numerically using Wang-Landau  sampling \cite{WangLandau}. An example plot of $S/N$ vs $E/N$ is shown in Fig \ref{SvsE} for a sub--community of 52 people.   We see that the entropy is very nearly a linear function of energy across a wide range of energies near the typical value.  It is not merely that $d^2 S/dE^2$ vanishes at a single critical point, but it is very nearly zero all together.    This unusual form of critical behavior was seen previously in the analysis of activity in networks of neurons \cite{BialekCriticality}.

\section{Coarse--graining social data}

The idea that social networks might be poised near a critical point is intriguing.  Related notions of criticality have emerged from the analysis of neural networks, but this has also generated controversy.  It is in principle possible that inference from finite data sets, using the maximum entropy framework, is biased toward finding models near criticality, or that some of the phenomenology which seems to be a signature of criticality could have more mundane explanations \cite{CriticalityInferredModel,Schwab,LatentVar,macke}.  One response to these concerns is to look very closely and ask whether the alternatives to criticality really explain the data in detail, as discussed for a population of neurons in Ref \cite{NeuronCriticality}.  But the approach to a critical point seems so dramatic that we should be able to give a more direct argument.

  \begin{figure}[t]
   \includegraphics[width=\linewidth]{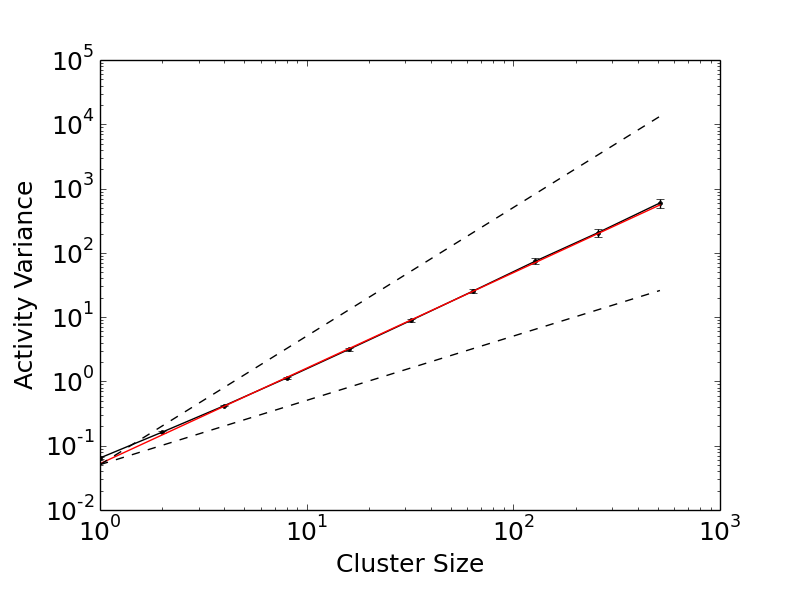} 
   \caption{\textbf{ Variance Scaling.} The variance of $\sigma _i ^{(k)}$ against cluster size $K = 2^k$. Dotted lines indicate linear and quadratic scaling; red line indicates a power--law fit with exponent $1.49 \pm .02$. Error bars are the standard deviation over random halves of the data. Fit obtained by nonlinear least squares with reduced $\chi^2  = 0.26$. \label{var-K}}
\end{figure}

In our modern view, a critical point can be defined as a nontrivial fixed point of the renormalization group (RG) \cite{Sethna, Fisher}.   In the standard formulation, microscopic variables live in real space, and have their dominant interactions with near neighbors.  The RG involves averaging over spatial neighborhoods \cite{Kadanoff}, and then tracking the distribution of these coarse--grained variables as a function of the averaging scale.    A crucial result is that the joint distributions of coarse--grained variables become simpler as the scale becomes larger, so that most interactions are ``irrelevant'';  models of macroscopic behavior thus are simpler and more universal than the underlying microscopic details.    More subtly, there are special parameter settings such that the joint distribution of coarse--grained variables is invariant to the scale of averaging. This is a fixed point of the RG transformation, and these fixed points correspond to critical points.

In order to explore RG ideas in more complex systems we have to find coarse--graining strategies that do not lean on the locality of interactions.  Recent work on large populations of neurons suggests that a natural analog of averaging with spatial neighbors is averaging with maximally correlated partners \cite{RGNeurons}, and we follow this approach here.  In brief, we walk through the network, identifying maximally correlated pairs $i,j_*(i)$, and then add the corresponding variables together,
\begin{equation}
\sigma_i ^{(2)} = \sigma _i ^{(1)} + \sigma _{j_*(i)} ^{(1)} ,
\end{equation}
where superscripts refer to the level of coarse-graining. We then repeat this for the next most correlated pair of people and so on, until the original $N$ variables have become $N/2$.  If we iterate this full procedure $k$ times, then we turn our data on $N$ people into data on $N/K$ clusters, each with $K = 2^k$ people in them.

If  correlations are weak, the central limit theorem drives  the activity of the clustered variables toward  the normal distribution as the clustering scale increases.  Near criticality, the self--similar structure of correlations should evade the central limit theorem, driving the distribution toward a non--Gaussian fixed point.  In the same way that the central limit theorem predicts a linear scaling (for example) of the variance with the number of variables that are being summed, the approach to nontrivial fixed points typically is associated with different scaling behavior.  It is this scaling behavior that we are looking for in the data.
 
The means of the coarse--grained variables scale linearly with the cluster size $K$,  so the first interesting question concerns the variance.  For independent variables the variance will be linear in the cluster size $K$, while for perfectly correlated variables the growth would be quadratic.  In Fig \ref{var-K} we show the behavior of the variance vs cluster size in  a community of 583 people, and we see that the variance has near perfect scaling at an intermediate exponent  $\sim 1.5$. This intermediate scaling indicates that there is nontrivial structure to the correlations in the dataset,  independent of the level of coarse--graining.   When comparing across different communities, we see a range of scaling exponents between 1.29 and 1.69. We could not discern any clear pattern or clustering in the scaling of different communities.   Although there is significant variability across communities, the precision of scaling within communities is surprising.

\begin{figure}[t]
   \includegraphics[width=\linewidth]{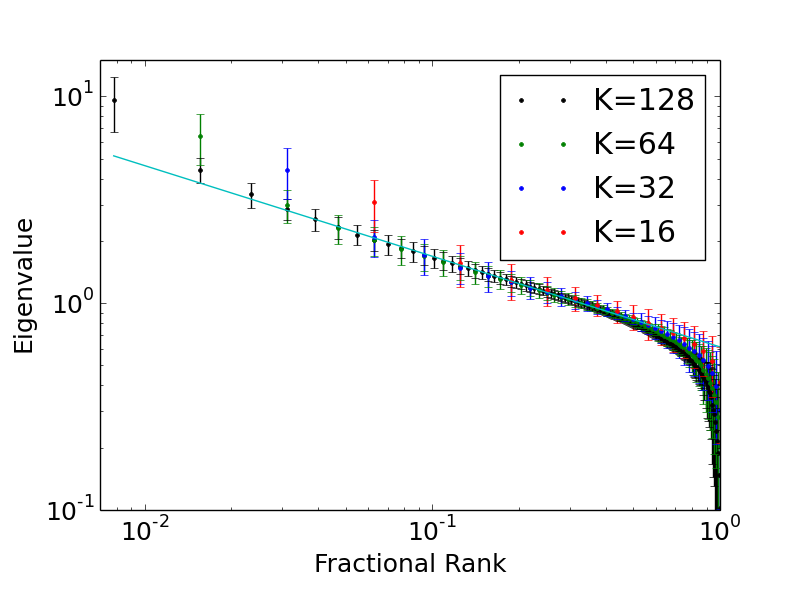} 
   \caption{\textbf{ Correlation Spectra.} Correlation spectra of various cluster sizes plotted as a function of the fractional rank. Error bars are standard deviations across different clusters and random halves of the data. Cyan line is a power--law  with exponent $-0.438 \pm 0.015$.  \label{spectrum}}
\end{figure}

We can also look at the structure of correlations within clusters. In the analogy with physical critical points, we  expect correlations to have the same structure at each stage of coarse--graining. In translation invariant systems with spatially local interactions, this corresponds to a scale-free correlation function, or equivalently a power--law dependence of the spectrum on momentum. In our systems, the closest analogue to this correlation function is the spectrum of the correlation matrix within clusters \cite{PCARG,RGNeurons}.    In Figure \ref{spectrum} we plot the averaged spectra of cluster correlation matrices, for different clusters sizes $K$, as a function of the fractional rank of the spectrum for the same community as used in Fig \ref{var-K}.  We stop at $K=128$ to avoid contamination of the spectrum by finite sample effects.

As we can see in Figure \ref{spectrum}, the  clusters seem to have a correlation spectrum that is independent of the size of the cluster. That is, plotted as a function of fractional rank, the spectra collapse onto a single curve that is independent of the degree of coarse-graining, with the exception of the largest eigenvalue of each cluster. Furthermore, these spectra seem to exhibit a power law dependence as a function of their fractional rank for a large range.  Although noisy, even this largest eigenvalue seems to have a regular behavior as a function of cluster size.  The correlations are scale free both in the sense that they are independent of the degree of coarse-graining and in the sense that they have a power law form.

\section{Discussion}

The dynamics of human behavior on social media are complex, and what we have done here is a first try.  Nonetheless, we find it striking that the social states of these networks have relatively simple, orderly behavior that can be captured in the language of statistical physics.  The joint distribution of activity in a community is described quite accurately by models that match only pairwise correlations, and are equivalent to familiar Ising models.  More deeply, the parameters of these models seem not to be arbitrary, but rather are poised near critical surfaces, and we see independent evidence of this near--criticality in the scaling behavior of the system under coarse--graining.

It is an old idea that complex systems, far from equilibrium, might organize themselves to states that are analogous to critical states in equilibrium statistical physics \cite{Bak}.  One can think of many reasons why such an organization might be advantageous:  the system becomes infinitely sensitive to (some) small signals, distant parts of the system can exchange information, the system grows long time scales, and more, although in different contexts the same features might be disadvantageous.  Importantly, all of these features arise together at the critical point, and so it is necessarily difficult to disentangle which ones are actually functional.  

There is an intuitive if  non-rigorous connection between criticality  and some familiar properties of social systems. Specifically, the high susceptibility and information transfer in critical networks evoke the tendency of online content to `go viral' or to very quickly spread over a long distance in a social network.  Just as small perturbations become amplified in critical networks, a social phenomenon initiated by a small group of people can quickly become amplified in online social networks.

While it is tempting to suggest that the proximity of a critical point is the ``mechanism'' by which things go viral on a social network,   it is difficult to imagine a mechanism for social systems to tune themselves to this kind of critical point. Generically, any system capable of producing such complex behavior will likely be controlled by many different parameters, and critical dynamics will only hold in a relatively small section of this high-dimensional  space. It is unclear   how an online social system would naturally tune itself to this area of its parameter space.   Nonetheless, this is what we see.
 
As far as we know, neither the maximum entropy method nor the renormalization group has been used previously in thinking about  social networks.    As the social science community accumulates more ``big data,'' more such tools will be needed.  We hope to have made clear that these relatively simple ideas, grounded in statistical physics,  are quite successful in revealing interesting regularities of human behavior in these social systems.

\begin{acknowledgments}
We thank Alex Aparicio, Curtis Callan, and Ned Wingreen   for  helpful  comments and discussions.  This work was supported in part by the National Science Foundation through the Center for the Physics of Biological Function (PHY--1734030), the Center for the Science of Information (CCF--0939370), and  PHY--1607612.
\end{acknowledgments}

\appendix
\section{Acquiring data}

\begin{figure}[t]
  \includegraphics[width=\columnwidth]{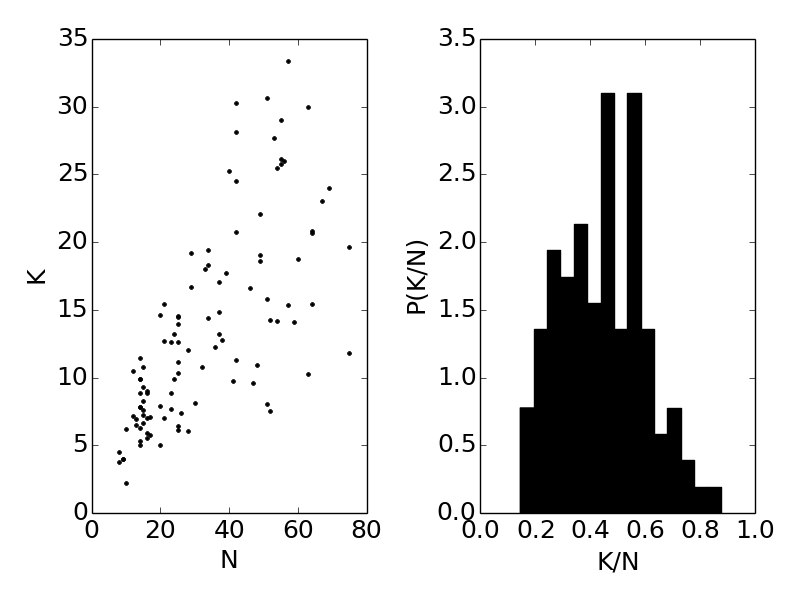} 
\caption{ \textbf{Topological Characteristics.} For 106 sub--communities with inferred pairwise max-ent models. (Left) Mean degree of social network $K$ against community size $N$ for all communities. (Right) Distribution of $K/N$. \label{topology}}
\end{figure}

Raw data were scraped from publicly available tweets using the Twitter API (https://developer.twitter.com/). As described in the main text, for each dataset a root user was chosen and a social network was built out to second degree from that root. That is, we find who the root user follows, and then who those people follow; our examples of social networks are built from those connections.  These networks were then reduced, identifying sub--communities using the Clauset-Newman-Moore algorithm \cite{community}.  The resulting sub--communities contain between 8 and 80 people, and vary considerably in topology, as summarized in Fig \ref{topology}.

Mean degree (at left in Fig \ref{topology}) represents the typical number of social connections that a given individual in a sub--community has within that sub--community. Below a sub--community size $N \sim 40$ people, the mean degree $K$ grows roughly linearly with the system size. For communities larger than 40 people, there is no discernible relationship between community size and the number of social connections. Interestingly, while one might imagine that these two types of social communities have different behaviors, in the communities that we have examined max-ent models are capable of describing both types of systems.

\section{Defining keywords}

We define keywords to be words that are used many times while staying localized in a short period of time. Of course, we then must define how many times a word must be used and how localized a word must be to be considered a keyword. We use the standard deviation in the times that a word is used to quantify the degree to which a word is localized in time. 
These two criteria (number of times used, standard deviation in time) define a two dimensional space in which we can place each word that is used in a dataset. Our task is to find cutoffs in this space to define a clear set of keywords.  Unfortunately, this is hard.  All datasets examined are approximately Zipfian \cite{Zipf}, which means that there is no clear scale for usage, making a non-arbitrary cutoff for the number of times a word is used quite difficult.

\begin{figure}[b]
  \includegraphics[width=80mm]{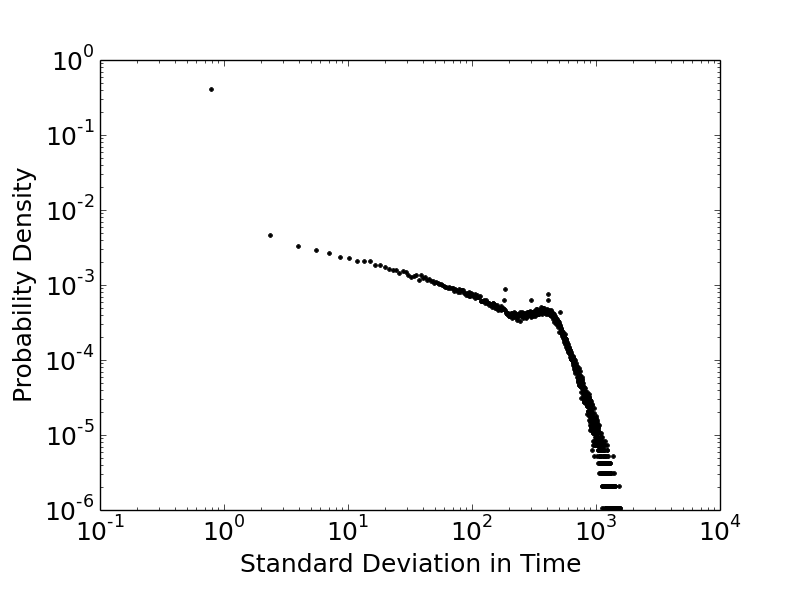} 
\caption{ \textbf{Distribution of Standard Deviations in Time of Word Usage.} For a dataset of 651 people. \label{stdDist}}
\end{figure}

The distribution of standard deviation in time for words is more interesting. We show the distribution of the standard deviation in time for language used in a dataset of 651 people in Fig \ref{stdDist}.  As we can see in Fig \ref{stdDist}, this distribution has two peaks, one corresponding to words that are used in a very short amount of time (far left) and words that are used with a standard deviation in time of around 500 days. The second peak begins with a kink in the distribution at a standard deviation of around 200 days. This second peak corresponds to words that are used independent of context, such as the staples of standard English vocabulary (`the', `she', etc).   Obviously, we do not want to include such words as keywords, as they are independent of context, and what makes keywords interesting is that they are highly contextual. We therefore bound the cutoff in standard deviation in time to be well clear of this second peak in Fig \ref{stdDist}. 

The statistics of word usage do not provide more detailed guidance for how to define keywords. As such, we examined the data at a range of different definitions for keywords \cite{gavThesis}, and generally found that the qualitative nature of results did not change with different definitions of keywords. For the purposes of presenting this data, we choose a definition for keywords that attempts to maximize the number of clearly meaningful data points. The exact parameters used for the datasets presented here are a cutoff in standard deviation in time of 130 days and a requirement that the word must be used at least 10 times in the data. In our datasets, this is generally sufficient to obtain on the order of 10 times the number of keywords as there are people in the dataset while still yielding generally comprehensible keywords.

\section{Maximum entropy models}

The maximum entropy method has a long history, and has received new attention in efforts to build statistical mechanics descriptions of biological networks directly from data.  Much of what we need thus is well known in some communities.  To make the discussion accessible to a wider community, we provide some review of these ideas here.    We start with geneal ideas and proceed to the specifics of our problem.

We recall that entropy, in addition to its thermodynamic meaning, provides the unique measure of available information consistent with simple and plausible conditions \cite{Shannon}.  Distributions with larger entropy thus describe variables about which we know less, a priori.  Maximizing the entropy is then a strategy for building models that inject as little structure or knowledge as possible, as first emphasized by Jaynes \cite{Jaynes,Jaynes2}.  More specifically, we want to insist that our models match certain features of the data, but otherwise have as little structure as possible.

In a system with states ${\boldsymbol \sigma}$, we can construct features $f_\mu({\boldsymbol \sigma})$, for $\mu =1 ,\, 2,\, \cdots ,\, K$.  Then if we are trying to make a model of the probability distribution $P({\boldsymbol \sigma})$, we insist that the average of these features in the model matches those seen experimentally,
\begin{equation}
\langle f_\mu ({\boldsymbol \sigma})\rangle_{\rm expt} = \sum_{\boldsymbol \sigma} P({\boldsymbol \sigma}) f_\mu ({\boldsymbol \sigma}) ,
\label{match1}
\end{equation}
for each $\mu$.  Among the distributions that obey this matching condition, we want to choose the one that maximizes the entropy
\begin{equation}
S[P(\boldsymbol \sigma )] = - \sum _{\boldsymbol \sigma} P(\boldsymbol \sigma) \log P (\boldsymbol \sigma) .
\label{entropy}
\end{equation}
To solve the constrained maximization problem we introduce Lagrange multipliers and define
\begin{widetext}
\begin{equation}
\tilde S [P(\boldsymbol \sigma)] =  -\sum _{\boldsymbol \sigma} P(\boldsymbol \sigma ) \log P(\boldsymbol \sigma)+ \\\sum_\mu \lambda _\mu \pn{\langle f_\mu ({\boldsymbol \sigma})\rangle_{\rm expt} - \sum _{\boldsymbol \sigma} f_\mu(\boldsymbol \sigma )P(\boldsymbol \sigma)} + \lambda _0 \pn{1- \sum _{\boldsymbol \sigma} P(\boldsymbol \sigma)},
\label{modEntropy2}
\end{equation}
\end{widetext}
where the Lagrange multipliers $\lambda_\mu$ correspond to each constrained feature and the term proportional to $\lambda_0$ constrains the distribution to be normalized.   Now we can search over all distributions, and adjust the values of the Lagrange multipliers at the end to be sure that the constraints are satisfied.

To maximize $\tilde S [P(\boldsymbol \sigma)]$, as usual we take the derivative and set it to zero,
\begin{equation}
\pp{\tilde S [P(\boldsymbol \sigma)]}{P(\boldsymbol \sigma)} = 0 ;
\label{diffEntropy}
\end{equation}
we can verify that the second derivatives are negative so that we really are finding a maximum of the entropy.  The solution is
\begin{equation}
P(\boldsymbol \sigma) = \fr{1}{Z} \exp \pn{-\sum _{\mu=1}^N \lambda _\mu f_\mu (\boldsymbol \sigma)} ,
\label{maxEntDist}
\end{equation}
where the partition function $Z$ absorbs $\lambda_0$ and thus depends implicitly on the values of all the other Lagrange multipliers,
\begin{equation}
Z = \sum_{\boldsymbol\sigma}  \exp \pn{-\sum _{\mu=1}^N \lambda _\mu f_\mu (\boldsymbol \sigma)} .
\label{Znorm}
\end{equation}

While Equation (\ref{maxEntDist}) gives the correct form of the max-ent model, it glosses over a major sticking point, which is that the correct values of the Lagrange multipliers $\lambda_\mu$ must be determined. This is a difficult computational problem \cite{InverseIsing}.

We use a method based on Monte Carlo sampling. Briefly, we simulate the model with some set of parameters $\{\lambda_\mu\}$ and then examine the expectation values of the observables $f_\mu $, computed as averages over the Monte Carlo samples in the simulation epoch labelled by $t$.   We then adjust the parameters by a factor proportional to the error  \cite{IsingRealNeurons, CollectiveNeurons}, so that the basic learning step is \begin{equation}
 \lambda _\mu (t+1) = \lambda _\mu (t) -\eta \left[  \expec{f_\mu}_t- \expec{f_\mu }_{\rm expt} \right] ,
\label{learningStep}
\end{equation}
where $\eta$ is a learning rate. For discussions about convergence see Refs \cite{DataDrivenAlg,MaxEntConvergence}.  We iterate until the differences between model and observed expectation values are comparable to the errors in the observed expectation values.

The expensive part of this procedure is generating new Monte Carlo samples for each set of parameters. To speed up this process we use the histogram Monte Carlo method, which allows us to recycle Monte Carlo samples generated with parameters $\boldsymbol \lambda $ to estimate features of the distribution parameterized by a different set of parameters $\boldsymbol \lambda '$ \cite{HistogramMC, MCFast}.  

In our particular case, the features that we choose are $f_\mu = \sigma_i$ and $f_\mu = \sigma_{i}\sigma_{j}$, for all values of $i$ and $j$ that share a social tie.  Constraining the expectation values of these features corresponds to fixing the probability that individual  $i$ participates in a Twitter conversation, and the correlations between participation by individuals $i$ and $j$.  With these choices, it is convenient to think of the Lagrange multipliers as ``effective fields'' $h_i$ and ``couplings" $J_{ij}$, and we arrive at the form of the model shown in Eqs (\ref{boltz}, \ref{E}) of the main text.  As indicated above, we need to follow the Monte Carlo procedure to arrive at values of these parameters given our measurements of $\langle \sigma_i \rangle$ and $\langle \sigma_i \sigma_j\rangle$.

It is useful to think about a null model in which we constrain only the probabilities that individuals  participate in a Twitter conversation, but ignore correlations.  Then the maximum entropy model has the same form as in Eqs (\ref{boltz}, \ref{E}), but with all $J_{ij} =0$ and 
\begin{equation}
h_i = \text{arctanh} (\langle \sigma _i \rangle) .
\label{indModel}
\end{equation}
This model is also equivalent to the assumption that each individual makes independent decisions about whether to tweet.

\section{Energy landscapes}

The model laid out in Eqs  (\ref{boltz}) and (\ref{E}) is similar to canonical models of spin glasses. Spin glasses tend to have many minima in their energy landscape due to frustration \cite{MezardSpinGlass}, which occurs when there are couplings of mixed sign in the Hamiltonian, competing with one another. This in turn leads to many metastable states in the energy landscape. 

\begin{figure}
  \includegraphics[width=80mm]{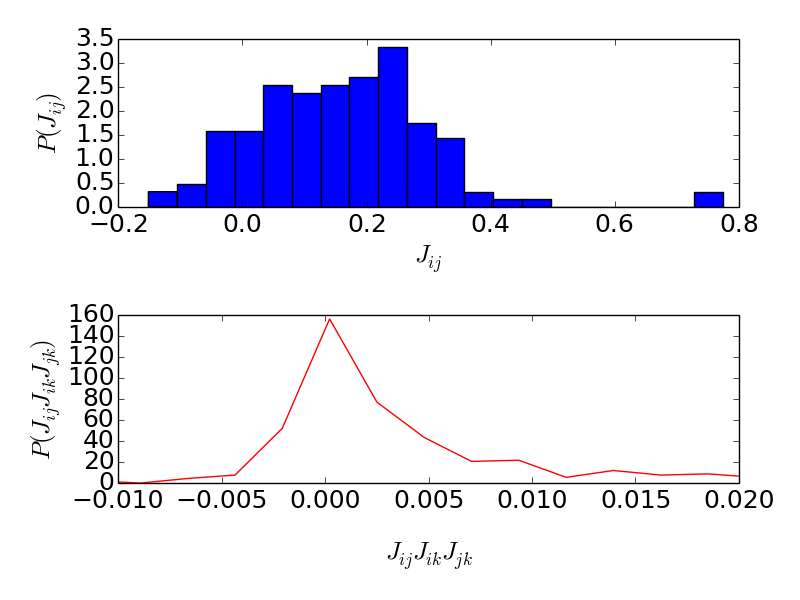} 
\caption{ \textbf{Characteristics of Coupling Matrix.} (Top) Distribution of couplings $J_{ij}$ for the community  in  Fig \ref{matrix}. (Bottom) Distribution of the product of all interactions from closed social triangles in the same community. Approximately 30\% of all possible triangles are frustrated. \label{frustration}}
\end{figure}

We can assess the complexity of the energy landscape by measuring how frequently frustration occurs. We examine this by looking at the distribution of the product of all coupling terms representing a social triangle (where three people all have social ties to one another) in the communities. When the product of these interaction terms is positive, there exists a social configuration that can minimize all relevant terms in the Hamiltonian. When this product is negative, then no state can minimize all relevant terms and the system is frustrated. We show an example of the distribution of couplings $J_{ij}$ and the distribution of the product of couplings from social triangles in Fig \ref{frustration}. As we can see, there are a significant number of frustrated triangles. This is true for all the communities that we examined. 

\begin{figure}[b]
  \includegraphics[width=80mm]{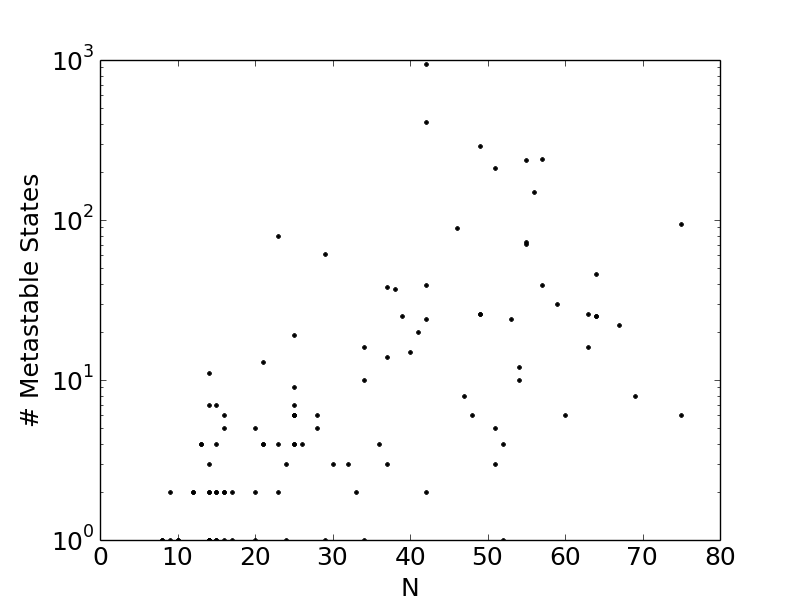} 
\caption{ \textbf{Number of Metastable States.} Number of metastable states, estimated as described in the text, for each of the communities that we analyzed. \label{basins}}
\end{figure}

We can also evaluate the complexity of the energy landscape by directly estimating the number of metastable states in the energy landscape, which we do by moving ``downhill'' in energy from each of $10^5$ Monte Carlo samples. We can then see how the number of metastable states scales with the size of the system. We show this relationship in Fig \ref{basins} for all 106 communities we examined.   Over the range that we can observe, the number of metastable states increases roughly exponentially with system size, albeit with considerable variation from instance to instance.  If this pattern persists into the thermodynamic limit it would put the energy landscape of these systems into the very complex class identified for the mean--field spin glass \cite{MezardSpinGlass}.

\section{Accuracy of three point correlations}

\begin{figure}
  \includegraphics[width=\columnwidth]{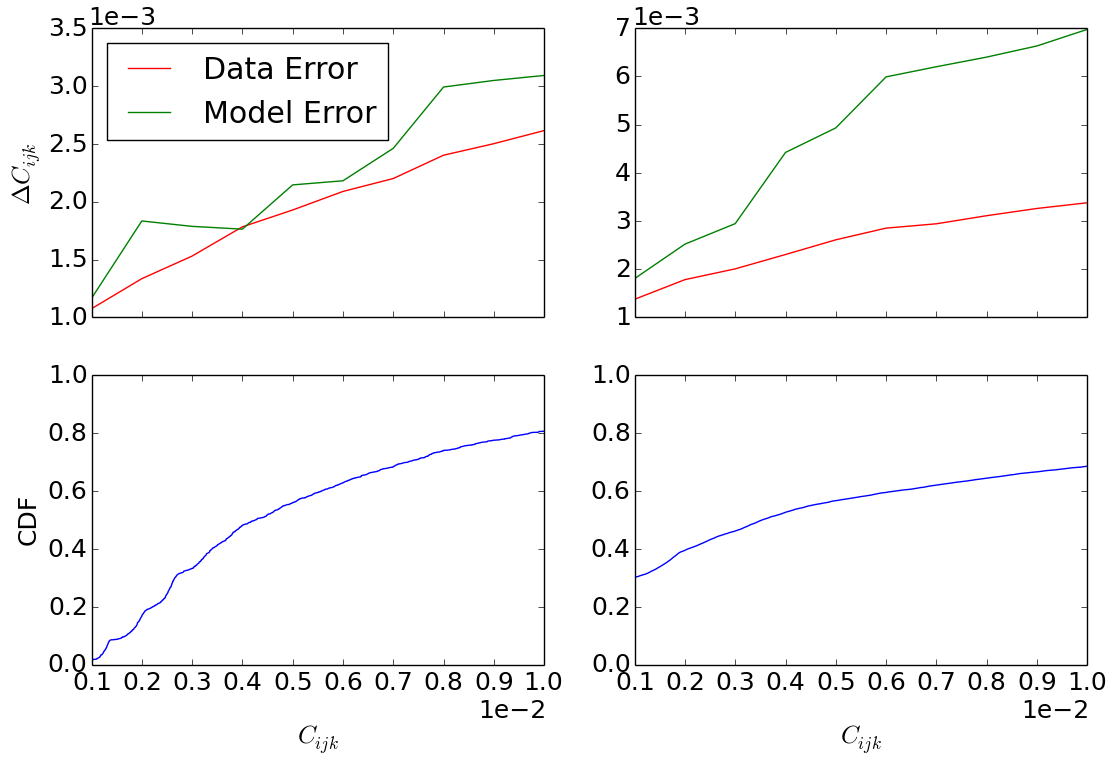} 
\caption{ \textbf{$C_{ijk}$ Prediction Error.} Top shows the root mean square error for predictions of three point correlations from the pairwise max-ent model (green) as well as the root mean square uncertainty for empirical three point correlations (red). Data were binned to compute root mean square errors. Bottom shows cumulative distribution of empirical three point correlation values. Left and right represent data from two different communities (data from left shown in Fig \ref{triplets}). \label{tripletError}}
\end{figure}

In Figure \ref{triplets} of the main text, each three point correlation $C_{ijk}$  carries an empirical uncertainty that can be estimated by bootstrap. The shear quantity of possible three point correlations in a heterogenous system makes it difficult to evaluate the accuracy of our predictions, so here we bin the three point correlations by their empirical values and then compare the root mean square error in prediction by the pairwise max-ent model to the root mean square uncertainty in the data.  In Figure \ref{tripletError} we show two examples of these errors. Data from the community on the left is shown in Fig  \ref{triplets} of  the main text.

For the community  on the left of Fig \ref{tripletError}, the prediction error is of the same order of magnitude as the measurement error, indicating that the pairwise max-ent model is able to capture three point correlations almost as well as the data allow. The bottom of Fig \ref{tripletError} shows the cumulative distribution of three point correlations. As we can see, the accuracy in the top of figure \ref{tripletError} covers the bulk of this distribution. However, for the community shown on the right of Fig \ref{tripletError}, the prediction error from the maximum entropy model is significantly larger than the intrinsic error in the data. In short, for the community on the right, the pairwise maximum entropy model is not capable of accurately reproducing 3 point correlations to the precision that the data allows. This does not mean that the maximum entropy model is incapable of making useful predictions on the three point correlations. Indeed, for the community on the right, the Pearson correlation coefficient between the empirical and predicted values of $C_{ijk}$ is $0.93$, which would normally be viewed as a success, even if we do not reach the maximum possible accuracy.

It is unclear what determines how well a pairwise model is capable of fitting data from a given community, and we suspect that variation between communities could be a fruitful area of study.

\section{Thermodynamics redux} 

If we can take our models seriously, then as the networks we study become larger the description in terms of statistical mechanics should imply an analog of thermodynamics.  We follow Refs \cite{NeuronCriticality,BialekCriticality,NaturalImages} in this construction, and for completeness we recall the arguments presented there.

The essential step is to write the partition function not as a sum over states but as an integral over the density of states,
\begin{equation}
Z = \sum _{\boldsymbol \sigma} e^{-E} = \int dE e^{-E}\rho (E),
\end{equation}
where
\begin{equation}
\rho(E) = \sum _{\boldsymbol \sigma } \delta (E- E_{\boldsymbol \sigma}).
\end{equation} 
We can integrate by parts to yield an expression in terms of the cumulative density of states, and from there the micro-canonical entropy [Eq (\ref{mc_ent})],
\begin{equation}
Z = \int dE e^ {S(E) - E} .
\end{equation}
We  then scale  the energy per node $\epsilon = E/N$ and the entropy per node $s = S(E)/N$, which gives us
\begin{equation}
Z = N \int d\epsilon  e^{N \pn {s(\epsilon) - \epsilon }} = \int dE e^{-Nf(\epsilon)},
\end{equation}
where $ f(\epsilon) =  \epsilon - s(\epsilon)$, is the free energy per particle.   The claim that
\begin{equation}
\lim_{N\rightarrow\infty} S(E)/N = s(\epsilon )
\end{equation}
is the claim that a thermodynamic limit exists, which is far from obvious.  But if it does exist, we can continue.

If we take the limit $N\to \ift$, we enter into the domain of Laplace's approximation, where $Z$ will be dominated by the minima of $f$. 
These minima are given by the solutions $\epsilon^*$ such that:
\begin{equation}
\dd{f}{\epsilon} \bigg | _{\epsilon ^ * } = 0 \imp \dd{s}{\epsilon} \bigg | _{\epsilon ^ * } = 1
\end{equation}
This is true for all systems, and is another way of defining temperature\cite{BialekCriticality,Kittel}, which we have set to be 1 in our discussion. 
Expanding to second order (as the first derivative disappears at the minima), we have that:
\begin{equation}
Z \approx N e^{-N f(\epsilon ^*)} \int d\epsilon \exp \pn{\fr{N}{2}  (\epsilon -\epsilon^* )^2 \ddd{s(\epsilon)}{\epsilon} \bigg | _{\epsilon ^*}}
\end{equation}
In this equation, it seems that the term outside the integral provides a contribution from the typical energy $\epsilon ^*$, while the term inside the integral bounds the  deviations from that typical energy. Crucially, the size of these deviations is controlled by the second derivative of the entropy with respect to the energy. When that second derivative is small, deviations from the typical energy will be large.   This is a critical point.

 \begin{figure}[b]
   \includegraphics[width=80mm]{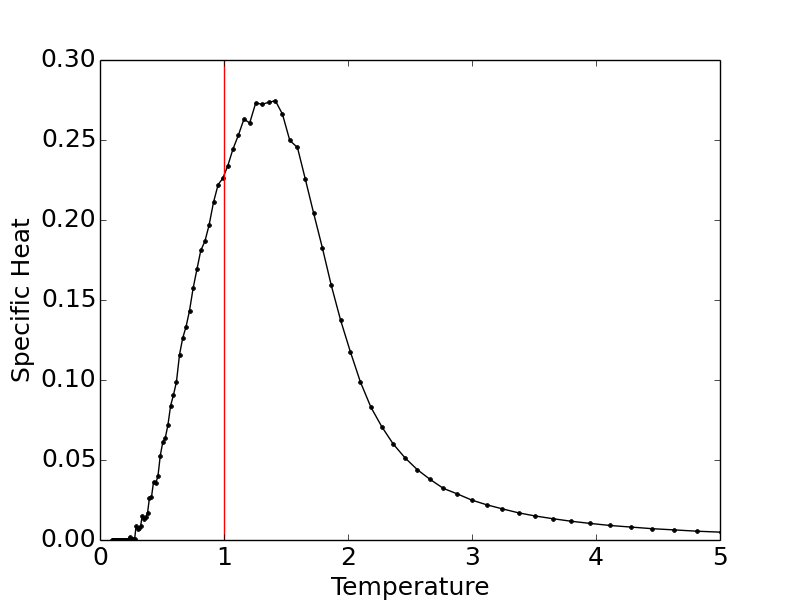} 
   \caption{\textbf{ Heat Capacity} Heat Capacity (equation \ref{specHeatEq}) as a function of a temperature coupled to the inferred max-ent Hamiltonian. Red Line indicates actual temperature ($T=1$). \label{specHeat}}
\end{figure}

The connection between the microcanonical entropy and the deviations from a system's typical energy is realized in the heat capacity, which can be expressed both in terms of the second derivative of the microcanonical entropy or in terms of the variance of the energy:
\begin{equation}
C_T = N \pn{ -\ddd{S}{E}}^{-1} = \sigma _E ^2 
\label{specHeatEq}
\end{equation}
Where $\sigma _E ^2$ is the variance of the energy.

A nearly linear entropy should correspond to a large value of the heat capacity. We can see this in Fig \ref{specHeat}, where we simulate the system shown in Fig \ref{chi} with a Hamiltonian scaled by various fictitious temperatures $H \to {H}/{T}$.   In Fig \ref{specHeat}, the real system (at $T = 1$) is slightly on the low temperature side of the peak in the heat capacity.  A peak in the heat capacity is another typical sign of criticality in physical systems, and it should increase our confidence that the systems examined here are near a critical point. 

\bibliography{paperBib}

\end{document}